\documentstyle[12pt,theorem]{ioplppt}

\theoremstyle{plain} \newtheorem{th}{Theorem}[section]
\theoremstyle{plain} \newtheorem{prop}[th]{Proposition}
\theoremstyle{plain} 
\theoremstyle{plain} \newtheorem{df}[th]{Definition}
{\theorembodyfont{\upshape} \theoremstyle{break}}

\newcount\N  \newcount\M  \newdimen\SIZE  \newdimen\INC
\def\YGBOX#1#2#3{
      \N=#1  \M=1  \INC=#2pt  \advance\INC by .#3pt  % Initialization
      \vbox{
         \loop\ifnum\M>0
            \M=\N
            \divide\N by 10         % Expansion of digits
            \multiply\N by 10
            \advance\M by -\N
            \divide\N by 10
            \SIZE=\INC
            \multiply\SIZE by \M    % Length of hruler
            \advance\SIZE by .#3pt
             \hrule  width \SIZE  height .#3pt
              \hbox{\loop\ifnum\M>0                      % Making of a row
                        \vrule  height #2pt  width .#3pt %  of Young diagram
                        \hskip #2pt
                        \advance\M by -1  \repeat
                        \vrule  width .#3pt }
             \hrule  width \SIZE  height .#3pt
            \vskip -.#3pt
         \repeat } }

\def\young#1{{             % Definition of command of Young diagram
       \mathchoice{\YGBOX{#1}61}{\YGBOX{#1}61}{\YGBOX{#1}41}{\YGBOX{#1}31}}}

\begin{document}
\jl{1}
%%%%%%%%%%%%%%%%%%% title %%%%%%%%%%%%%%%%%%%%%%%%%
\title{A generalization of determinant formulas for the solutions of 
       Painlev\'e II and XXXIV equations}[Determinant formulas for P$_{\rm II}$ and P$_{\rm XXXIV}$]

\author{Kenji Kajiwara\dag~and Tetsu Masuda\ddag}

\address{\dag\ Department of Electrical Engineering, \\
               Doshisha University, Kyotanabe, Kyoto, 610-0321 Japan}

\address{\ddag\ Department of Physics, Ritsumeikan University, 
                Kusatsu, Shiga, 525-8577 Japan}

%%%%%%%%%%%%%%% abstract %%%%%%%%%%%%%%%
\begin{abstract}
A generalization of determinant formulas for the classical solutions of
Painlev\'e XXXIV and Painlev\'e II equations are constructed using the
technique of Darboux transformation and Hirota's bilinear formalism. It
is shown that the solutions admit determinant formulas even for the
transcendental case.
\end{abstract}

%\maketitle

%%%%%%%%%%%%%%%%%%%%%%%%%%%%%%%%%%%%%%%%%%%%%%%%%%%%%%%%%%%%%%%%%%%%%%
\section{Introduction}
It is well-known that solutions for the Painlev\'e equations play a role
of special functions in the nonlinear science~\cite{Book}. Originally,
Painlev\'e derived these equations in order to find new transcendental
functions determined by second order ordinary differential equations
possessing so-called the Painlev\'e property. Recently, irreducibility of
solutions of the Painlev\'e equations has been proved by Umemura
\etal~\cite{Umemura:p4-p2,Umemura:100years}. Umemura first gave a
rigorous definition of classical functions: starting from the field of
rational functions, if a function is obtained by finite numbers of 
iterations of the following permissible operations, 
\begin{itemize}
 \item differentiation,
 \item arithmetic calculations,
 \item solving homogeneous linear ordinary differential equation,
 \item substitution into Abelian functions,
\end{itemize}
then that function is called ``classical''.

Umemura proved that solutions of the Painlev\'e equations are not
classical in general in the above sense. However, it is known that they
admit classical solutions for special values of parameters. One is the
rational or algebraic solutions, and another is the transcendental
classical solutions which are expressed by rational functions in various
special functions and their derivatives.

In this article, we discuss the Painlev\'e II equation (P$_{\rm II}$),
\begin{equation}
\frac{d^2u}{dz^2}=2u^3 - 4zu + 4\left(\alpha+\frac{1}{2}\right), \label{p2}
\end{equation}
and the equation of No. 34 in Gambier's classification~\cite{Ince}, 
\begin{equation}
2w\frac{d^2w}{dz^2}-\left(\frac{dw}{dz}\right)^2 +4w^3 -8zw^2 +16\alpha^2=0, \label{p34}
\end{equation}
which we call the Painlev\'e XXXIV equation (P$_{\rm XXXIV}$). 
In equations (\ref{p2}) and (\ref{p34}), 
we adopted different scale from their canonical forms~\cite{Ince}, 
\begin{equation}
\frac{d^2\tilde{u}}{ds^2}=2\tilde{u}^3+s\tilde{u} +\beta, \label{p2:canonical}
\end{equation}
\begin{equation}
2\tilde{w}\frac{d^2\tilde{w}}{ds^2}-\left(\frac{d\tilde{w}}{ds}\right)^2 -8\gamma\tilde{w}^3 +2s\tilde{w}^2 +1=0,  \label{p34:canonical}
\end{equation}
for the clarity of expression of solutions. 
The scale transformations, 
\begin{equation}
 \left.
  \begin{array}{@{\,}ll}
   \displaystyle
    s=(-4)^{1/3}z, \quad \tilde{u}=(-4)^{-1/3}u, \quad \tilde{w}=\pm2^{-4/3}\alpha^{-1}w,\\
   \displaystyle
    \beta=-\left(\alpha+\frac{1}{2}\right), \quad \gamma=\mp \alpha/2,
  \end{array}
 \right.
\end{equation}
lead the canonical forms to our ones. 
We denote
$u(\alpha)$, P$_{\rm II}[\alpha]$ and $w(\alpha)$, P$_{\rm XXXIV}[\alpha]$, respectively,
when it is necessary to show the value of parameter explicitly. 
These equations are related to
each other by the Miura transformation~\cite{Ince},
\begin{equation}
w=-\frac{du}{dz}-u^2+2z\ ,\label{miura}
\end{equation}
and its complement,
\begin{equation}
u=\frac{\displaystyle \frac{dw}{dz}+4\alpha}{2w}. \label{Imiura}
\end{equation}
Umemura \etal also proved that P$_{\rm II}$ (and thus P$_{\rm XXXIV}$) admits transcendental classical solutions
which are expressed by the Airy function when $\alpha$ is an integer, rational solutions when $\alpha$ is
a half-integer, and that otherwise the solutions are non-classical.

It is known that the Painlev\'e equations (except for P$_{\rm I}$) admit the
B\"acklund transformations(BT) which form the affine Weyl
group~\cite{Okamoto1,Okamoto2,Okamoto3,Okamoto4}. For example, BT of P$_{\rm II}$ is given by,
\begin{eqnarray}
S: S(u)=u+\frac{4\alpha}{\displaystyle \frac{du}{dz}+u^2-2z},
\quad S(\alpha)=-\alpha, \label{BT:S}\\
T: T(u)=-u+\frac{4(\alpha+1)}{\displaystyle \frac{du}{dz}-u^2+2z},
\quad T(\alpha)=\alpha+1,\label{BT:T}
\end{eqnarray}
and $<S,T>$ forms the affine Weyl group of type $A^{(1)}_1$. 
Starting from a suitable ``seed'' solution, we obtain ``higher'' solutions by
applying BT to it, which are expressed by rational
functions in the seed solution and its derivatives. 

It is interesting and important to note two points on classical
solutions for the Painlev\'e equations. The first is that, in general,
the classical solutions are located on special points in the parameter
space from the viewpoint of symmetry. Namely, transcendental classical
solutions are on the walls of the Weyl chambers, and rational solutions
on their barycenters.  The second is that the classical solutions have
additional structure. Namely, it is known that some of the classical solutions
admit such determinant formulas that they are expressed by log
derivative of the ratio of some determinants. Moreover, application of
BT corresponds to increment of size of the determinant. In fact,
both Airy function type and rational solutions for P$_{\rm II}$ and P$_{\rm XXXIV}$
admit such determinant structure.

Then, does such determinant structure exist even for non-classical
solutions, or are non-classical solutions so ``transcendental'' that
do not admit even such structure? Here we also note that
such determinant structure are shown to be universal among the soliton 
equations due to the celebrated Sato theory~\cite{Sato,elementary}.

The purpose of this article is to generalize the determinant formulas of
the classical solutions of P$_{\rm II}$ and P$_{\rm XXXIV}$, and show
that such determinant structure is universal among the solutions of
P$_{\rm II}$ and P$_{\rm XXXIV}$.  The key to this generalization is to
use the technique of the Darboux transformation together with Hirota's
bilinear formalism.

This article is organized as follows. In section \ref{classical}, we
summarize the derivation of P$_{\rm XXXIV}$ and classical solutions of
P$_{\rm II}$ and P$_{\rm XXXIV}$. We illustrate our basic ideas in
section \ref{Darboux}.  Our main results are presented in section
\ref{result}. The proof for our results are given in section
\ref{proof}. Section \ref{conc} is devoted to summary and discussions.

%%%%%%%%%%%%%%%%%%%%%%%%%%%%%%%%%%%%%%%%%%%%%%%%%%%%%%%%%%%%%%%%%%%%%%
\section{P$_{\rm XXXIV}$ and classical solutions of P$_{\rm II}$ and P$_{\rm XXXIV}$
\label{classical}}
It is well-known that P$_{\rm II}$ is derived as the similarity reduction 
from the modified KdV equation~\cite{Ablowitz},
\begin{equation}
u_{t}+\frac{3}{2}u^2u_{x}-\frac{1}{4}u_{xxx}=0.\label{mKdV}
\end{equation}
In fact, putting
\begin{equation}
u(x,t)=\frac{1}{(3t)^{1/3}}u(z),\quad z=\frac{x}{(3t)^{1/3}},
\end{equation}
equation (\ref{mKdV}) is reduced to equation (\ref{p2}). 
Similarly, P$_{\rm XXXIV}$ (\ref{p34}) is derived 
as the similarity reduction from the KdV equation,
\begin{equation}
w_t-\frac{3}{2}ww_x-\frac{1}{4}w_{xxx}=0,\label{KdV}
\end{equation}
by putting
\begin{equation}
w(x,t)=\frac{1}{(3t)^{2/3}}(w(z)-2z),\quad z=\frac{x}{(3t)^{1/3}}.
\end{equation}

It is known that P$_{\rm II}$ (\ref{p2}) is bilinearized to~\cite{p2:rational}
\begin{eqnarray}
\left(D_z^2-\mu\right)g\cdot f=0,\label{p2:bl1}  \\
\left[D_z^3+(4z-3\mu)D_z-4\left(\alpha+\frac{1}{2}\right)\right]g\cdot f=0,\label{p2:bl2}
\end{eqnarray}
by the dependent variable transformation,
\begin{equation}
u=\frac{d}{dz}\log\frac{g}{f},
\end{equation}
where $\mu$ is an arbitrary function in $z$, and $D_z$ is Hirota's
bilinear differential operator defined by
\begin{equation}
D_z^n~g \cdot f=\left.\left(\frac{d}{dz}-\frac{d}{dz'}\right)^ng(z)f(z')\right|_{z'=z}.
\end{equation}
The bilinear equations (\ref{p2:bl1}) and (\ref{p2:bl2}) are regarded as
those for P$_{\rm II}$, but it is also possible to derive P$_{\rm
XXXIV}$ as follows. First we divide equations (\ref{p2:bl1}) and
(\ref{p2:bl2}) by $f^2$. Then, putting
\begin{equation}
\psi=\frac{g}{f},\quad w=2\frac{d^2}{dz^2}\log f -\mu+2z,\label{dep:p34}
\end{equation}
and using the formulas~\cite{Network,Hirota:Book},
\numparts
\begin{eqnarray}
\frac{D_z g\cdot f}{f^2}=\frac{d}{dz}\psi,  \\
\frac{D_z^2 g\cdot f}{f^2}=\left(\frac{d^2}{dz^2}+w+\mu-2z\right)\psi,  \\
\frac{D_z^3 g\cdot f}{f^2}=\left[\frac{d^3}{dz^3}+3(w+\mu-2z)\frac{d}{dz}\right]\psi,
\end{eqnarray}
\endnumparts
we obtain
\begin{eqnarray}
\left( 2w\frac{d}{dz}-\frac{dw}{dz}-4\alpha\right)\psi = 0, \label{lin1:p34}  \\
\left(\frac{d^2}{dz^2}+w-2z\right)\psi=0.                   \label{lin2:p34}
\end{eqnarray}
Compatibility condition of the linear equations (\ref{lin1:p34}) and (\ref{lin2:p34}) 
gives P$_{\rm XXXIV}[\alpha]$.

Let us discuss the determinant expressions for the classical solutions of P$_{\rm II}$ and P$_{\rm
XXXIV}$ ~\cite{Okamoto3,p2:rational,akira}. In the following, the determinant with size zero should
be regarded as $1$.

If we choose $\mu=2z$, then it is known that the bilinear equations
(\ref{p2:bl1}) and (\ref{p2:bl2}) are satisfied by 
\begin{equation}
g=\rho_{N+1},\quad f=\rho_N, \quad \alpha=N,\quad N\in {\bf Z}_{\geq 0},
\end{equation}
\begin{equation}
\rho_N=
 \left|
  \begin{array}{cccc}
   f^{(0)}   & f^{(1)} & \cdots & f^{(N-1)}  \\
   f^{(1)}   & f^{(2)} & \cdots & f^{(N)}    \\
   \vdots    & \vdots  & \ddots & \vdots     \\
   f^{(N-1)} & f^{(N)} & \cdots & f^{(2N-2)} 
  \end{array}
 \right|,    \quad \rho_0=1,
\end{equation}
where $\displaystyle f^{(m)}=\frac{d^m}{dz^m}f$ and $f$ satisfies the Airy equation, 
\begin{equation}
\frac{d^2f}{dz^2}=2zf\ .\label{airy}
\end{equation}
Thus, the Airy function type solutions of P$_{\rm II}[N]$ and
P$_{\rm XXXIV}[N]$ are given by 
\numparts
\begin{eqnarray}
u=\frac{d}{dz}\log\frac{\rho_{N+1}}{\rho_N}, \\
w=2\frac{d^2}{dz^2}\log\rho_N, 
\end{eqnarray}
\endnumparts
respectively. 

The rational solutions are obtained for the case of $\mu=0$, 
and there are two expressions for them. 
One is the Schur function type expression given by 
\begin{equation}
g=\sigma_{N+1},\quad f=\sigma_N,\quad \alpha=N+\frac{1}{2},\quad N\in {\bf Z}_{\geq 0},
\end{equation}
\begin{equation}
\sigma_N=
 \left|
  \begin{array}{cccc}
   q_N      & q_{N+1}  & \cdots & q_{2N-1} \\
   q_{N-2}  & q_{N-1}  & \cdots & q_{2N-3} \\
   \vdots   & \vdots   & \ddots & \vdots   \\
   q_{-N+2} & q_{-N+3} & \cdots & q_1
  \end{array}
 \right|,    \quad \sigma_0=1,
\end{equation}
where $q_k$ are the polynomials in $z$ defined by 
\begin{equation}
\sum_{k=0}^\infty q_k(z)~\xi^k=\exp\left(z\xi+\frac{\xi^3}{3}\right),
\quad q_k(z)=0\ \mbox{for}\ k<0\ .
\end{equation}
Another expression is given by Hankel determinant as 
\begin{equation}
g=\kappa_{N+1},\quad f=\kappa_{N},\quad \alpha=N+\frac{1}{2},\quad N\in {\bf Z}_{\geq 0},
\end{equation}
\begin{equation}
\kappa_N = 
 \left|
  \begin{array}{cccc}
   a_0    & a_1    & \cdots & a_{N-1}  \\
   a_1    & a_2    & \cdots & a_N      \\
   \vdots & \vdots & \ddots & \vdots   \\
   a_{N-1}& a_N    & \cdots & a_{2N-2}
  \end{array}
 \right| ,   \quad \kappa_0=1, \label{Hankel}
\end{equation}
where $a_n$, $n=0,1,2,\cdots$ are the polynomials in $z$, which are
recursively defined by
\begin{equation}
a_{n} = \frac{da_{n-1}}{dz} + \sum_{k=0}^{n-2}a_ka_{n-k-2},
\quad n> 0,\quad a_0=z.
\label{orecursion}
\end{equation}
Thus, the rational solutions for P$_{\rm II}[N+\frac{1}{2}]$ and P$_{\rm
XXXIV}[N+\frac{1}{2}]$  are given by 
\numparts
\begin{eqnarray}
u=\frac{d}{dz}\log\frac{\sigma_{N+1}}{\sigma_N}
 =\frac{d}{dz}\log\frac{\kappa_{N+1}}{\kappa_N}, \\
w=2z+2\frac{d^2}{dz^2}\log\sigma_N
 =2z+2\frac{d^2}{dz^2}\log\kappa_N,
\end{eqnarray}
\endnumparts
respectively. 

We note that it is possible to generalize the above result for negative $\alpha$
by the reflection symmetry of P$_{\rm II}$ and P$_{\rm XXXIV}$,
\begin{equation}
 u(-\alpha-1)=-u(\alpha),\label{sym:u}
\end{equation}
and
\begin{equation}
w(-\alpha)=w(\alpha),\label{sym:w}
\end{equation}
respectively. Indeed, this symmetry is generated by $ST$ in equations (\ref{BT:S}) and (\ref{BT:T}).

%%%%%%%%%%%%%%%%%%%%%%%%%%%%%%%%%%%%%%%%%%%%%%%%%%%%%%%%%%%%%%%%%%%%%%
\section{Darboux Transformation\label{Darboux}}
The technique of Darboux transformation is well developed in the soliton
theory~\cite{Matveev}. For example, in the case of KdV equation
(\ref{KdV}), we start with its auxiliary linear problem,
\begin{equation}
 \left.
  \begin{array}{@{\,}ll}
   \displaystyle
    -\Psi_{xx}-w\Psi=\lambda\Psi, \\
   \displaystyle
    \Psi_t=\Psi_{xxx}+\frac{3}{2}w\Psi_x+\frac{3}{4}w_x\Psi,
  \end{array}
 \right.     \label{lin:KdV}
\end{equation}
where $\lambda$ is the spectral parameter. Then, one can show that
equations (\ref{lin:KdV}) are covariant with respect to the Darboux
transformation $\Psi \to \Psi[N]$, $w \to w[N]$ defined by
\begin{equation}
\Psi[N]=\frac{Wr(\Psi_1,\Psi_2,\cdots,\Psi_N,\Psi)}{Wr(\Psi_1,\Psi_2,\cdots,\Psi_N)},
\end{equation}
\begin{equation}
w[N]=w+2\frac{\partial^2}{\partial x^2}\log Wr(\Psi_1,\Psi_2,\cdots,\Psi_N),
\end{equation}
where $\Psi_k$ is a solution of the linear equations (\ref{lin:KdV})
with $\lambda=\lambda_k$, and $Wr$ is a Wronskian with respect to the
indicated functions. Thus, choosing an appropriate seed solution of
KdV equation as $w$, one can construct series of exact solutions by
using this method. For example, starting from the solution $w=0$, then
we obtain the Wronskian expression of $N$-soliton
solution~\cite{Satsuma}. Although most of the
expression would be somewhat formal except for the soliton and rational
type solutions or their variants, it is important that we can express quite
wide class of solutions in terms of determinant.

The above result is recovered by the following bilinear equations:
\begin{equation}
 \left( D_xD_t -\frac{1}{4}D_x^4-\frac{3}{2}wD_x \right)F\cdot F=0,
\end{equation}
\begin{eqnarray}
&&\left(D_x^2+\lambda+w\right)G\cdot F=0,\label{g-bl1:KdV}\\
&& \left[D_x^3-4D_t+3(-\lambda+w)D_x\right]G\cdot F=0,
\label{g-bl2:KdV}
\end{eqnarray}
where $F=Wr(\Psi_1,\Psi_2,\cdots,\Psi_N)$,
$G=Wr(\Psi_1,\Psi_2,\cdots,\Psi_N,\Psi)$.
We omit the detail, but we can directly prove that these bilinear
equations holds as the identity of the determinants from the linear
equations (\ref{lin:KdV}).
Indeed, we recover ``usual'' bilinear equations by putting $w=0$,
\begin{equation}
 \left( D_xD_t -\frac{1}{4}D_x^4\right)F\cdot F=0,
\end{equation}
\begin{eqnarray}
&&\left(D_x^2+\lambda\right) G\cdot F=0,\label{bl1:KdV}\\
&& \left(D_x^3-4D_t-3\lambda D_x\right)G\cdot F=0.
\label{bl2:KdV}
\end{eqnarray}
>From these bilinear equations, we recover the KdV equation (\ref{KdV})
and the auxiliary linear problem (\ref{lin:KdV}) by putting as
\begin{equation}
 w=2\frac{\partial^2}{\partial x^2}\log F,
\quad \Psi=\frac{G}{F}.
\end{equation}

Now, keeping this correspondence between the Darboux transformation and
the bilinear formalism in mind, let us illustrate our strategy to
generalize the determinant formulas for the solutions of P$_{\rm II}$
and P$_{\rm XXXIV}$. We have the bilinear equations
(\ref{p2:bl1}) and (\ref{p2:bl2}) for P$_{\rm XXXIV}$.
Clearly, these bilinear equations
correspond to equations (\ref{bl1:KdV}) and (\ref{bl2:KdV}). Thus, if we could
obtain ``generalized'' bilinear equations which correspond to
equations (\ref{g-bl1:KdV}) and (\ref{g-bl2:KdV}), it might be possible to
find the Darboux transformation for P$_{\rm XXXIV}$ and P$_{\rm II}$
which leaves the linear equations (\ref{lin1:p34}) and (\ref{lin2:p34})
covariant. Then we obtain a generalization of the determinant formulas.

In the next section, we present our main results.

\section{Main Results\label{result}}
In this section, we present a generalization of determinant formulas 
for the solutions of P$_{\rm II}$ and P$_{\rm XXXIV}$. 

We start with a pair of linear equations (\ref{lin1:p34}) and
(\ref{lin2:p34}). As we mentioned in section \ref{classical}, these
equations are compatible provided that $w$ satisfies P$_{\rm
XXXIV}[\alpha]$. In other words, equation (\ref{lin2:p34}) follows if $w$ and
$\psi$ are solutions of P$_{\rm XXXIV}[\alpha]$ and equation (\ref{lin1:p34}),
respectively. Then we have the following theorem.

\begin{th}\label{main1}
Let $w$ be a solution of P$_{\rm XXXIV}[\alpha]$, 
and $\psi_0$ be a solution of the linear equation (\ref{lin1:p34}). 
We define two sequences $\psi_n$, $\varphi_n$ ($n=0,1,2,\cdots$) by
\begin{eqnarray}
&&\varphi_0=\frac{w}{2\psi_0},\label{rec0:p34}\\
&& \psi_n = \frac{d\psi_{n-1}}{dz}
+\frac{w}{2\psi_0}\sum_{k=0}^{n-2}\psi_k\psi_{n-2-k},\quad n>0,\label{rec1:p34}\\
&&\varphi_n =
\frac{d\varphi_{n-1}}{dz}+\psi_0\sum_{k=0}^{n-2}\varphi_k\varphi_{n-2-k},\quad n>0.\label{rec2:p34}
\end{eqnarray}
We define Hankel determinant $\tau_N$, $N\in {\bf Z}$ by
\begin{equation}
\tau_N = \left\{
  \begin{array}{cc}
\det(\psi_{i+j-2})_{i,j=1,\cdots N}& N>0\\
1 & N=0\\
\det(\varphi_{i+j-2})_{i,j=1,\cdots -N}& N<0\\
\end{array}
\right. .\label{tau:p34}
\end{equation}
Then, 
\begin{equation}
\Psi_N = \frac{\tau_{N+1}}{\tau_N}\label{wave:p34}
\end{equation}
satisfies 
\begin{eqnarray}
\left[ 2W\frac{d}{dz}-\frac{dW}{dz}-4(\alpha+N)\right]\Psi_N = 0, \label{lin3:p34}\\
\left(\frac{d^2}{dz^2}+W-2z\right)\Psi_N=0,                       \label{lin4:p34}
\end{eqnarray}
where
\begin{equation}
W = w + 2\frac{d^2}{dz^2}\log\tau_N.\label{w:p34}
\end{equation}
Thus, $W$ satisfies P$_{\rm XXXIV}[\alpha+N]$.
\end{th}
Similar formula for P$_{\rm II}$ is obtained by applying the Miura
transformation (\ref{miura}).  The linear equations (\ref{lin1:p34}) and
(\ref{lin2:p34}) are reduced to
\begin{equation}
\left(\frac{d}{dz}-u\right)\psi=0.  \label{lin:p2}
\end{equation}
Then we have the following theorem. 
\begin{th}\label{main2}
Let $u$ be a solution of P$_{\rm II}[\alpha]$, 
and $\tau_N$ be Hankel determinant given by equation (\ref{tau:p34}), 
where $\psi_n$ and $\varphi_n$ are defined recursively by 
\begin{equation}
\psi_0=\exp \left( \int dz~u \right),\quad
\varphi_0=\frac{\displaystyle -\frac{du}{dz}-u^2+2z}{2\psi_0},\label{rec0:p2}
\end{equation}
 \begin{equation}
\psi_n = \frac{d\psi_{n-1}}{dz}
+\varphi_0\sum_{k=0}^{n-2}\psi_k\psi_{n-2-k},
\quad n>0, \label{rec1:p2}
 \end{equation}
\begin{equation}
\varphi_n = \frac{d\varphi_{n-1}}{dz}
+\psi_0\sum_{k=0}^{n-2}\varphi_k\varphi_{n-2-k},
\quad n>0, \label{rec2:p2}
\end{equation}
Then, 
\begin{equation}
U = \frac{d}{dz}\log\frac{\tau_{N+1}}{\tau_N}\label{u:p2}
\end{equation}
satisfies P$_{\rm II}[\alpha+N]$. 
\end{th}
As mentioned in the previous section, Theorems \ref{main1} and
\ref{main2} is the direct consequence of the following proposition:
\begin{prop}\label{bl}
The following bilinear equations hold. 
\begin{eqnarray}
(D_z^2+w-2z)\tau_{N+1}\cdot\tau_N=0,\label{bl1} \\
\left[D_z^3+(3w-2z)D_z-4\left(\alpha+N+\frac{1}{2}\right)\right]
\tau_{N+1}\cdot\tau_N=0.\label{bl2}
\end{eqnarray}
\end{prop}
In fact, dividing equations (\ref{bl1}) and (\ref{bl2}) by $\tau_N^2$ 
and using the formula~\cite{Network,Hirota:Book}
\numparts
\begin{eqnarray}
\frac{D_z\tau_{N+1}\cdot\tau_N}{\tau_N^2}=\frac{d}{dz}\Psi_N,  \\
\frac{D_z^2\tau_{N+1}\cdot\tau_N}{\tau_N^2}=\left(\frac{d^2}{dz^2}+W-w\right)\Psi_N,  \\
\frac{D_z^3\tau_{N+1}\cdot\tau_N}{\tau_N^2}
=\left[\frac{d^3}{dz^3}+3(W-w)\frac{d}{dz}\right]\Psi_N,
\end{eqnarray}
\endnumparts 
we obtain the linear equations (\ref{lin3:p34}) and
(\ref{lin4:p34}) in Theorem \ref{main1}. Similarly, dividing 
equations (\ref{bl1}) and (\ref{bl2}) by $\tau_{N+1}\tau_N$ and using
the formulas~\cite{Network,Hirota:Book}, 
\numparts
\begin{eqnarray}
\frac{D_z   \tau_{N+1} \cdot \tau_N}{\tau_{N+1}\tau_N} = U,  \\
\frac{D_z^2 \tau_{N+1} \cdot \tau_N}{\tau_{N+1}\tau_N} = V+U^2,  \\
\frac{D_z^3 \tau_{N+1} \cdot \tau_N}{\tau_{N+1}\tau_N} = \frac{d^2 U}{dz^2} +3UV+U^3,
\end{eqnarray}
\endnumparts
with
\begin{equation}
V = \frac{d^2}{dz^2}\log \left( \tau_{N+1}\tau_N \right), 
\end{equation}
we get 
\begin{eqnarray}
V+U^2+w-2z=0,  \\
\frac{d^2U}{dz^2}+3UV+(3w-2z)U-4\left(\alpha+N+\frac{1}{2}\right)=0. 
\end{eqnarray}
Eliminating $V$, we obtain 
\begin{equation}
\frac{d^2U}{dz^2}=2U^3 -4zU +4\left(\alpha+N+\frac{1}{2}\right). 
\end{equation}

These results imply that even non-classical solutions possess the same
determinant structure as the classical solutions. Indeed, the known
determinant expressions for the classical solutions are recovered as
special cases. In fact, starting with a solution $w=2z$ for P$_{\rm XXXIV}[\frac{1}{2}]$, 
the linear equation (\ref{lin1:p34}) yields
\begin{equation}
\left(z\frac{d}{dz}-1\right)\psi = 0,
\end{equation}
from which we get $\psi_0=z$ without losing generality. 
Then, we have from the recursion relations (\ref{rec0:p34}), (\ref{rec1:p34}) and (\ref{rec2:p34}),
\begin{equation}
\psi_n = \frac{d\psi_{n-1}}{dz}+\sum_{k=0}^{n-2}\psi_k\psi_{n-2-k},
\quad n>0,\quad \psi_0=z,
\label{rec:rational:p34}
\end{equation}
\begin{equation}
\varphi_n = \frac{d\varphi_{n-1}}{dz}+z\sum_{k=0}^{n-2}\varphi_k\varphi_{n-2-k},
\quad n>0,\quad \varphi_0=1.
\label{rec2:rational:p34}
\end{equation}
Thus, we have a series of rational solutions given by equations
(\ref{tau:p34}), (\ref{rec:rational:p34}), (\ref{rec2:rational:p34}) and  
\numparts
\begin{eqnarray}
U=\frac{d}{dz}\log\frac{\tau_{N+1}}{\tau_N}, \\
W=2z+2\frac{d^2}{dz^2}\log\tau_N.
\end{eqnarray}
\endnumparts
The case of $N\geq 0$ agrees with the Hankel determinant expression for
the rational solutions discussed in section \ref{classical}.

Next, noticing that $w=0$ is a solution of P$_{\rm XXXIV}[0]$, 
we find from equation (\ref{lin2:p34}) that $\psi_0$ is determined by
\begin{equation}
\left(\frac{d^2}{dz^2}-2z\right)\psi=0,  \label{airy2}
\end{equation}
and the recursion relation (\ref{rec1:p34}) with equation (\ref{rec0:p34}) is reduced to 
\begin{equation}
\psi_n = \frac{d\psi_{n-1}}{dz}.  \label{rec:airy}
\end{equation}
Thus, we have a series of solutions given by equations
(\ref{tau:p34}), (\ref{airy2}), (\ref{rec:airy}) and 
\numparts
\begin{eqnarray}
U=\frac{d}{dz}\log\frac{\tau_{N+1}}{\tau_N}, \\
W=2\frac{d^2}{dz^2}\log\tau_N,
\end{eqnarray}
\endnumparts 
for $N\geq 0$, which agrees with the Wronskian
expression for the Airy function type solutions discussed in section
\ref{classical}. However, the Airy function type solutions are ``singular'' case
in our formula. From equations (\ref{rec0:p34}) and (\ref{rec2:p34}), we find that $\varphi_n=0$ for all $n$ and
thus $\tau_N=0$ for $N<0$, which does not give meaningful result. We note that this phenomena is
related to the symmetry of the $\tau$ sequence which will be mentioned in section \ref{conc}, and
the correct result for $N<0$ case is obtained by virtue of symmetry. We also note that
this is the only singular case and our determinant formula works for other cases.

%%%%%%%%%%%%%%%%%%%%%%%%%%%%%%%%%%%%%%%%%%%%%%%%%%%%%%%%%%%%%%%%%%%%%%
\section{Proof of theorems \label{proof}}
In this section, we give the proof of the Theorems \ref{main1} and
\ref{main2}. Since we see that these theorems follows
immediately from Proposition \ref{bl}, it is sufficient to prove it.
We have to prove both $N>0$ and $N<0$ cases, but we
concentrate on the former case, since the latter case is proved in
similar manner.

The bilinear equations (\ref{bl1}) and (\ref{bl2}) are reduced to the 
Pl\"ucker relations, which are quadratic identities of the determinants
whose columns are shifted.
Thus, we first construct such differential formulas that shifted
determinants are expressed by operating some differential operator
on the original determinant. For this purpose, 
we introduce the following notation:
\begin{df}
Let $Y$ be a Young diagram $Y=(i_1,i_2,\cdots,i_h)$. 
Then we define an $N\times N$ determinant $\tau_{NY}$ by 
\begin{equation}
\fl
\tau_{NY} = 
 \left|
  \begin{array}{cccccccc}
   \psi_0 & \psi_1 & \cdots & \psi_{N-h-1} 
          & \psi_{N-h+i_h}  & \cdots & \psi_{N-2+i_2} & \psi_{N-1+i_1} \\
   \psi_1 & \psi_2 & \cdots & \psi_{N-h} 
          & \psi_{N-h+1+i_h} &\cdots & \psi_{N-1+i_2} &\psi_{N+i_1}    \\
   \vdots & \vdots & \cdots & \vdots    & \vdots & \cdots & \vdots & \vdots     \\
   \psi_{N-1} & \psi_N    &\cdots & \psi_{2N-h-2} 
              & \psi_{2N-h-1+i_h} & \cdots & \psi_{2N-3+i_2}& \psi_{2N-2+i_1}
  \end{array}
 \right|.
\end{equation}
\end{df}
Then, we have the following differential formulas.
\begin{prop}\label{shift}
\begin{eqnarray}
\hspace*{-40pt}
\tau_N{}_{\young{1}}=\frac{d}{dz} \tau_N, \label{shift1} \\
\hspace*{-40pt}
\tau_N{}_{\young{2}}+\tau_N{}_{\young{11}}
=\left(\frac{d^2}{dz^2}+\frac{w}{2}\right) \tau_N, \label{shift2} \\
\hspace*{-40pt}
\tau_N{}_{\young{2}}-\tau_N{}_{\young{11}}
=\left(-\frac{w}{2}+2Nz\right)\tau_N\ ,   \label{shift3}  \\
\hspace*{-40pt}
\tau_N{}_{\young{3}}+2\tau_N{}_{\young{12}}+\tau_N{}_{\young{111}}
=\left(\frac{d^3}{dz^3}+\frac{3}{2}w\frac{d}{dz}+\frac{1}{2}\frac{dw}{dz}\right)
\tau_N, \label{shift4}\\
\hspace*{-40pt}
\tau_N{}_{\young{3}}-\tau_N{}_{\young{111}}
=\left[ \left(-\frac{w}{2}+2Nz\right)\frac{d}{dz}
+2\left(\alpha-1+N\right)\right]\tau_N,\label{shift5}\\
\hspace*{-40pt}\tau_N{}_{\young{3}}-\tau_N{}_{\young{12}}+\tau_N{}_{\young{111}}
=\left[2z\frac{d}{dz}+2N^2+4N(\alpha-1) 
-\frac{1}{4}\frac{dw}{dz}-3(\alpha-1)\right]\tau_N.\label{shift6}
\end{eqnarray}
\end{prop}
The proof for Proposition \ref{shift} is an important step.
However, since this requires straightforward but tedious calculations,
we will give it in \ref{proof-shift}. 

Finally, we prove Proposition \ref{bl}. 
>From the Pl\"ucker relations we have, 
\begin{equation}
\tau_{N+1}{}_{\young{11}}\tau_N
-\tau_{N+1}{}_{\young{1}}\tau_N{}_{\young{1}}
+\tau_{N+1}\tau_N{}_{\young{2}}=0,\label{pl1}
\end{equation}
\begin{equation}
\tau_{N+1}{}_{\young{12}}\tau_N
-\tau_{N+1}{}_{\young{2}}\tau_N{}_{\young{1}}
+\tau_{N+1}\tau_N{}_{\young{3}}=0.\label{pl2}
\end{equation}
By using Proposition \ref{shift}, we get the bilinear relations
(\ref{bl1}) and (\ref{bl2}) from equations (\ref{pl1}) and (\ref{pl2}),
respectively, which is the desired result.
%%%%%%%%%%%%%%%%%%%%%%%%%%%%%%%%%%%%%%%%%%%%%%%%%%%%%%%%%%%%%%%%%%%%%%
\section{Summary and Discussions\label{conc}}
In this article, we have presented determinant formulas for the solutions of 
P$_{\rm XXXIV}$ and P$_{\rm II}$ which are valid also for non-classical
solutions by using the technique of the Darboux transformation and bilinear formalism.
The solutions of P$_{\rm XXXIV}[\alpha+N]$ and P$_{\rm II}[\alpha+N]$, 
$N\in {\bf Z}$, are expressed by determinants
whose entries are constructed from the solution of some linear equations.
Moreover, coefficients of those linear equations include the solution of 
P$_{\rm XXXIV}[\alpha]$ and P$_{\rm II}[\alpha]$, respectively. 
We have also shown that known determinant expressions 
for classical solutions are recovered as special cases. This result implies
that determinant structure of the classical solutions is universal among the solutions
of P$_{\rm II}$ and P$_{\rm XXXIV}$.

Finally, let us discuss the relation with the Toda equation, which
will be a key for generalization to other Painlev\'e equations.

In general, the $\tau$ function for P$_{\rm II}$ is introduced through
its Hamiltonian~\cite{Okamoto3},
\[
 H_{\rm II}(v,u,z;\alpha)= \frac{1}{2}v^2 + (-u^2+2z)v - 4\alpha u,
\]
by
\begin{equation}
 H_{\rm II}(v,u,z;\alpha)=\frac{d}{dz}\log \tau(\alpha).
\end{equation}
We note that we obtain P$_{\rm II}[\alpha]$ for $u$ from the canonical
equation,
\begin{equation}
\frac{du}{dz}=\frac{\partial H_{\rm II}}{\partial v},\quad
\frac{dv}{dz}=-\frac{\partial H_{\rm II}}{\partial u}.
\end{equation}
Then it can be shown that $u(\alpha)$, which is a solution of
P$_{\rm II}[\alpha]$, is expressed as
\begin{equation}
u(\alpha)=\frac{d}{dz}\log \frac{\tau(\alpha+1)}{\tau(\alpha)}.
\end{equation}
By applying BT, we obtain a sequence of $\tau$ functions
$\{\tau_N\}_{N\in{\bf Z}}$,
where $\tau_N=\tau(\alpha+N)$. Okamoto has shown that BT of P$_{\rm II}$
is governed by the Toda equation on the level of $\tau$ function:
\begin{prop}(Okamoto~\cite{Okamoto3})
$\tau_N$ satisfies the Toda equation,
\begin{equation}
\frac{d^2}{dz^2}\log\tau_N = c_N\frac{\tau_{N+1}\tau_{N-1}}{\tau_N^2},\label{Toda}
\end{equation}
where $c_N$ is a non-zero constant.
\end{prop}
It is well-known that the solution of the Toda equation (\ref{Toda})
is expressed by 
\begin{equation}
\tau_N=\det\left(\frac{d^{i+j-2}}{dz^{i+j-2}}f\right)_{i,j=1,\cdots N},
\label{Toda:Darboux}
\end{equation}
where $f$ is an arbitrary function in $z$. Equation (\ref{Toda:Darboux}) 
is sometimes referred as ``Darboux's formula''. However, one
should note that Darboux's formula (\ref{Toda:Darboux}) is valid under
the condition,
\begin{equation}
\tau_0=1,\quad \tau_1=f,\quad N\geq 0\label{bc:Darboux}
\end{equation}
(In practice, $\tau_0$ can be a constant.) 
In \cite{Okamoto3}, it is pointed out that the Wronskian expression
for the Airy function type solutions of P$_{\rm II}$ is a consequence of 
the Darboux's formula. We demonstrate how we could apply the 
Darboux's formula for this case. We have a solution of P$_{\rm II}$ for $\alpha=0$,
\begin{equation}
 u(0)=\frac{d}{dz}\log \psi,\quad \frac{d^2}{dz^2}\psi=2z\psi.
\end{equation}
Therefore, we can choose the $\tau$ functions as
\begin{equation}
 \tau(0)=\tau_0=1,\quad \tau(1)=\tau_1=\psi.
\end{equation}
Then we have a $\tau$ sequence for the Airy function
type solutions,
\begin{equation}
 \cdots\tau_{-3},\quad \tau_{-2},\quad
\tau_{-1},\quad \tau_0=\tau(0)=1,\quad \tau_1=\psi,\quad \tau_2,\quad\tau_3\cdots .
\end{equation}
Now, as was mentioned in section \ref{classical},
we have a reflection symmetry (\ref{sym:u}) on $u$ 
which implies a symmetry on the $\tau$ function,
\begin{equation}
 \tau(-\alpha)=\tau(\alpha).\label{sym:tau}
\end{equation}
In the case of the Airy function type solutions, this symmetry induces
a symmetry on the $\tau$ sequence as
\begin{equation}
 \tau_{-N}=\tau_N.
\end{equation}
By virtue of this symmetry, we see that the $\tau$ sequence is
divided into two parts as
\begin{equation}
 \cdots\tau_{3},\quad \tau_{2},\quad
\tau_{1},\quad \tau_0=\tau(0)=1,\quad \tau_1=\psi,\quad \tau_2,\quad\tau_3\cdots .
\end{equation}
Fortunately enough, since it can be shown that this $\tau$ sequence is
governed by the Toda equation (\ref{Toda}) with $c_N=1$, we could apply
Darboux's formula for $N>0$ and $N<0$ separately. 
We note that if we prolong the $\tau$ sequence for $N<0$ 
following to the Toda equation (\ref{Toda}) with the condition (\ref{bc:Darboux}) without taking the symmetry
into account, we have $\tau_N=0$ for $N<0$. This corresponds to the ``singular'' phenomena mentioned
in section \ref{result}.

In the case of rational solutions, the $\tau$ sequence is again separated into two
parts. However, we cannot apply the Darboux's formula to this case. 
Let us take a solution of P$_{\rm II}$ for $\alpha=1/2$,
\begin{equation}
 u\left(\frac{1}{2}\right)=\frac{1}{z}.
\end{equation}
We can choose the $\tau$ function as
\begin{equation}
 \tau\left(\frac{1}{2}\right)=\tau_0=1,\quad
\tau\left(\frac{3}{2}\right)=\tau_1=z,
\end{equation}
and we have a $\tau$ sequence for the rational solutions. Now the
symmetry (\ref{sym:tau}) implies a symmetry on the $\tau$ sequence,
\begin{equation}
 \tau_N=\tau_{-N-1},\label{sym:C-Toda}
\end{equation}
and thus we have,
\begin{equation}
 \cdots\tau_{2},\quad \tau_{1},\quad
\tau_{0},\quad \tau_0=\tau\left(\frac{1}{2}\right)=1,
\quad \tau_1=z,\quad \tau_2,\quad\tau_3\cdots,
\end{equation}
and again the $\tau$ sequence is separated into two parts.
However in this case, these $\tau$ functions are shown to 
satisfy the Toda equation of the form,
\begin{equation}
\frac{d^2}{dz^2}\log \tau_N = \frac{\tau_{N+1}\tau_{N-1}}{\tau_N^2}-z.
\end{equation}
Thus, we cannot apply Darboux's formula for the rational solutions.
Of course, by introducing a gauge on the $\tau$ function as
\begin{equation}
\sigma_N={\rm e}^{z^3/6}~\tau_N,
\end{equation}
then $\sigma_N$ satisfy the Toda equation
\begin{equation}
\frac{d^2}{dz^2}\log \sigma_N = \frac{\sigma_{N+1}\sigma_{N-1}}{\sigma_N^2}.
\end{equation}
However, now $\sigma_0$ is not a constant. 

There are two points for being able to apply the Darboux's formula.  The
first is that $\tau$ sequence should have a symmetry which is induced
from the Painlev\'e equation itself so that the $\tau$ sequence is
separated into two parts, which is necessary to apply the Darboux's
formula without inconsistency. The second is that $\tau$ sequence
should satisfy the Toda equation of the form (\ref{Toda}) under
the condition (\ref{bc:Darboux}).  Both conditions are satisfied for the
Airy function type solution , but the second condition does not
hold for rational solutions. 
Now, for solutions which corresponds to generic $\alpha$, the symmetry on the $\tau$ function
(\ref{sym:tau}) does not induce any symmetry on the $\tau$ sequence.
This observation shows that it is not trivial that generic
$\tau$ function admit determinant formula, even if it satisfies the Toda
equation.

Despite of unavailability of the Darboux's formula, 
we could obtain the determinant formula for
rational solutions. This is due to the determinant formula of the
general solution of Toda equation of C-type
(Toda equation with the symmetry (\ref{sym:C-Toda})) obtained 
in~\cite{p2:rational}.

Conversely, the general determinant formula for P$_{\rm II}$
strongly implies that it is possible to construct the determinant
formula for the general solution of Toda equation in general
setting, i.e., the Toda equation admits the determinant formula for the general solution without
any symmetries or gauge on $\tau$ sequence. Then, our results might be
regarded as the special case of such general solution for the Toda
equation.

Moreover, it is known that the B\"acklund transformations for
the Painlev\'e equations (except for P$_{\rm I}$) is governed by
various types of Toda equations~\cite{Okamoto1,Okamoto2,Okamoto3,Okamoto4}. Thus, it might be possible to
present general determinant formula also for other Painlev\'e equations.
We shall work out this problem in the next publication.

%%%%%%%%%%%%%%%%%%%%%%%%%%%%%%%%%%%%%%%%%%%%%%%%%%%%%%%%%%%%%%%%%%%%%%
\ack 
The authors would like to thank Dr. Y. Ohta for valuable
suggestions. They also thank Prof. H. Umemura, Prof. M. Noumi and
Prof. Y. Yamada for interest to our work and discussions.  One of the
authors (K K) is supported by the Grant-in-aid for Encouragement of
Young Scientists, The Ministry of Education, Science and Culture of
Japan, No. 09740164.

%%%%%%%%%%%%%%%%%%%%%%%%%%%%%%%%%%%%%%%%%%%%%%%%%%%%%%%%%%%%%%%%%%%%%%
\appendix
\section{\label{proof-shift}}
Here, we give the proof for Proposition~\ref{shift}. 

We first prove equation (\ref{shift1}). Notice that $\tau_N{}_{\young{1}}$ is
expressed by
\begin{equation}
\hspace*{-25pt}
\tau_N{}_{\young{1}} = 
 \left(
  \begin{array}{cccc}
   \psi_1 & \psi_2     & \cdots & \psi_N      \\
   \psi_2 & \psi_3     & \cdots & \psi_{N+1}  \\
   \vdots & \vdots     & \ddots & \vdots      \\
   \psi_N & \psi_{N+1} & \cdots & \psi_{2N-1}
  \end{array}
  \right)    \cdot
 \left(
  \begin{array}{cccc}
   \Delta_{11} & \Delta_{12} & \cdots & \Delta_{1N}  \\
   \Delta_{21} & \Delta_{22} & \cdots & \Delta_{2N}  \\
   \vdots      & \vdots      & \ddots & \vdots       \\
   \Delta_{N1} & \Delta_{N2} & \cdots & \Delta_{NN}
  \end{array}
  \right),   \label{shift1-pro}
\end{equation}
where $\Delta_{ij}$ is the $(i,j)$-cofactor of $\tau_N$ and $A \cdot B$
denotes a standard scalar product for $N \times N$ matrices $A=(a_{ij})$
and $B=(b_{ij})$ which is defined as
\begin{equation}
A \cdot B = \sum_{i,j=1}^N a_{ij}b_{ij} = {\rm trace}~A~^tB.
\end{equation}
The first matrix of equation (\ref{shift1-pro}) is rewritten by using the
recursion relation (\ref{rec1:p34}) as
\begin{eqnarray}
\hspace*{-20pt}
\left(
 \begin{array}{cccc}
  \displaystyle \frac{d}{dz}\psi_0 & \displaystyle \frac{d}{dz}\psi_1 
                                   & \cdots & \displaystyle \frac{d}{dz}\psi_{N-1}  \\
  \displaystyle \frac{d}{dz}\psi_1 & \displaystyle \frac{d}{dz}\psi_2 
                                   & \cdots & \displaystyle \frac{d}{dz}\psi_N      \\
  \vdots             & \vdots                 & \ddots & \vdots                     \\
  \displaystyle \frac{d}{dz}\psi_{N-1} & \displaystyle \frac{d}{dz}\psi_N 
                                   & \cdots & \displaystyle \frac{d}{dz}\psi_{2N-2}
 \end{array}
\right)     \nonumber  \\
\hspace*{-20pt}
+ \frac{w}{2\psi_0}
\left(
 \begin{array}{cccc}
  0 & \psi_0^2 & \cdots & \displaystyle \sum_{k=0}^{N-2}\psi_k \psi_{N-2-k}  \\
  \psi_0^2 & \psi_0\psi_1+\psi_1\psi_0 & \cdots 
                                       & \displaystyle \sum_{k=0}^{N-1}\psi_k \psi_{N-1-k}  \\
  \vdots & \vdots & \ddots & \vdots  \\
  \displaystyle \sum_{k=0}^{N-2}\psi_k \psi_{N-2-k} & 
  \displaystyle \sum_{k=0}^{N-1}\psi_k \psi_{N-1-k} & \cdots & 
  \displaystyle \sum_{k=0}^{2N-3}\psi_k \psi_{2N-3-k}
 \end{array}
\right).
\end{eqnarray}
The above matrix in the second term is separated as 
\begin{eqnarray}
\left(
 \begin{array}{cccc}
  0 & 0 & \cdots & 0  \\
  \psi_0^2 & \psi_1\psi_0 & \cdots & \psi_{N-1} \psi_0  \\
  \vdots & \vdots & \ddots & \vdots  \\
  \displaystyle \sum_{k=0}^{N-2}\psi_k \psi_{N-2-k} & 
  \displaystyle \sum_{k=1}^{N-1}\psi_k \psi_{N-1-k} & \cdots & 
  \displaystyle \sum_{k=N-1}^{2N-3}\psi_k \psi_{2N-3-k}
 \end{array}
\right)     \nonumber \\
+
\left(
 \begin{array}{cccc}
  0 & \psi_0^2          & \cdots & \displaystyle \sum_{k=0}^{N-2}\psi_k \psi_{N-2-k}  \\
  0 & \psi_0 \psi_1     & \cdots & \displaystyle \sum_{k=0}^{N-2}\psi_k \psi_{N-1-k}  \\
  \vdots & \vdots       & \ddots & \vdots  \\
  0 & \psi_0 \psi_{N-1} & \cdots & \displaystyle \sum_{k=0}^{N-2}\psi_k \psi_{2N-3-k}
 \end{array}
\right)
\end{eqnarray}
Each of these terms gives zero contribution in equation (\ref{shift1-pro}). 
Hence we obtain equation (\ref{shift1}). 

Next we prove equation (\ref{shift2}). We consider 
\begin{equation}
\fl
\tau_N{}_{\young{2}}+\tau_N{}_{\young{11}} = 
 \left(
  \begin{array}{ccccc}
   \psi_1 & \psi_2     & \cdots & \psi_{N-1}  & \psi_{N+1}  \\
   \psi_2 & \psi_3     & \cdots & \psi_N      & \psi_{N+2}  \\
   \vdots & \vdots     & \ddots & \vdots      & \vdots      \\
   \psi_N & \psi_{N+1} & \cdots & \psi_{2N-2} & \psi_{2N}
  \end{array}
  \right)    \cdot
 \left(
  \begin{array}{cccc}
   \Delta{}_{\young{1}}{}_{11} & \Delta{}_{\young{1}}{}_{12} & \cdots 
                               & \Delta{}_{\young{1}}{}_{1N}  \\
   \Delta{}_{\young{1}}{}_{21} & \Delta{}_{\young{1}}{}_{22} & \cdots 
                               & \Delta{}_{\young{1}}{}_{2N}  \\
   \vdots      & \vdots        & \ddots & \vdots       \\
   \Delta{}_{\young{1}}{}_{N1} & \Delta{}_{\young{1}}{}_{N2} & \cdots 
                               & \Delta{}_{\young{1}}{}_{NN}
  \end{array}
  \right),
\end{equation}
where $\Delta{}_{\young{1}}{}_{ij}$ is $(i,j)$-cofactor of $\tau_N{}_{\young{1}}$. 
The first matrix in the right-hand side is equal to 
\begin{eqnarray}
\fl
\left(
 \begin{array}{ccccc}
  \displaystyle \frac{d}{dz}\psi_0 & \displaystyle \frac{d}{dz}\psi_1 & \cdots 
& \displaystyle \frac{d}{dz}\psi_{N-2}  & \displaystyle \frac{d}{dz}\psi_N          \\
  \displaystyle \frac{d}{dz}\psi_1 & \displaystyle \frac{d}{dz}\psi_2 & \cdots 
& \displaystyle \frac{d}{dz}\psi_{N-1}  & \displaystyle \frac{d}{dz}\psi_{N+1}       \\
  \vdots             & \vdots           & \ddots & \vdots                            \\
  \displaystyle \frac{d}{dz}\psi_{N-1}  & \displaystyle \frac{d}{dz}\psi_N & \cdots 
& \displaystyle \frac{d}{dz}\psi_{2N-3} & \displaystyle \frac{d}{dz}\psi_{2N-1}
 \end{array}
\right)     \nonumber  \\
\fl
+ \frac{w}{2\psi_0}
\left(
 \begin{array}{ccccc}
  0 & 0 & \cdots & 0 & 0  \\
  \psi_0^2 & \psi_1\psi_0 & \cdots & \psi_{N-2} \psi_0 & \psi_N \psi_0  \\
  \vdots & \vdots & \ddots & \vdots & \vdots  \\
  \displaystyle \sum_{k=0}^{N-2}\psi_k \psi_{N-2-k} & 
  \displaystyle \sum_{k=1}^{N-1}\psi_k \psi_{N-1-k} & \cdots & 
  \displaystyle \sum_{k=N-2}^{2N-4}\psi_k \psi_{2N-4-k} & 
  \displaystyle \sum_{k=N}^{2N-2}\psi_k \psi_{2N-2-k}
 \end{array}
\right)     \nonumber  \\
\fl
+ \frac{w}{2\psi_0}
\left(
 \begin{array}{ccccc}
  0 & \psi_0^2          & \cdots & \displaystyle \sum_{k=0}^{N-3}\psi_k \psi_{N-3-k} 
                        & \displaystyle \sum_{k=0}^{N-1}\psi_k \psi_{N-1-k}         \\
  0 & \psi_0 \psi_1     & \cdots & \displaystyle \sum_{k=0}^{N-3}\psi_k \psi_{N-2-k} 
                        & \displaystyle \sum_{k=0}^{N-1}\psi_k \psi_{N-k}           \\
  \vdots & \vdots       & \ddots & \vdots & \vdots  \\
  0 & \psi_0 \psi_{N-1} & \cdots & \displaystyle \sum_{k=0}^{N-3}\psi_k \psi_{2N-4-k} 
                        & \displaystyle \sum_{k=0}^{N-1}\psi_k \psi_{2N-2-k}
 \end{array}
\right).
\end{eqnarray}
Taking the scalar product, the first and second terms give
$\displaystyle \frac{d}{dz}\tau_N{}_{\young{1}}$ and $\displaystyle
\frac{w}{2}\tau_N$, respectively, and the third term vanishes. Hence we
have equation (\ref{shift2}).

Next we prove equation (\ref{shift3}). We consider the following equality, 
\begin{equation}
\fl
\tau_N{}_{\young{2}}-\tau_N{}_{\young{11}} = 
 \left(
  \begin{array}{cccc}
   \psi_2     & \psi_3     & \cdots & \psi_{N+1}  \\
   \psi_3     & \psi_4     & \cdots & \psi_{N+2}  \\
   \vdots     & \vdots     & \ddots & \vdots      \\
   \psi_{N+1} & \psi_{N+2} & \cdots & \psi_{2N}
  \end{array}
  \right)    \cdot
 \left(
  \begin{array}{cccc}
   \Delta_{11} & \Delta_{12} & \cdots & \Delta_{1N}  \\
   \Delta_{21} & \Delta_{22} & \cdots & \Delta_{2N}  \\
   \vdots      & \vdots      & \ddots & \vdots       \\
   \Delta_{N1} & \Delta_{N2} & \cdots & \Delta_{NN}
  \end{array}
  \right).
\end{equation}
The first matrix of right-hand side is rewritten as 
\begin{eqnarray}
\left(
 \begin{array}{cccc}
  \displaystyle \frac{d}{dz}\psi_1 & \displaystyle \frac{d}{dz}\psi_2 & \cdots 
                                   & \displaystyle \frac{d}{dz}\psi_N           \\
  \displaystyle \frac{d}{dz}\psi_2 & \displaystyle \frac{d}{dz}\psi_3 & \cdots 
                                   & \displaystyle \frac{d}{dz}\psi_{N+1}       \\
  \vdots             & \vdots           & \ddots & \vdots                       \\
  \displaystyle \frac{d}{dz}\psi_N & \displaystyle \frac{d}{dz}\psi_{N+1} & \cdots 
                                   & \displaystyle \frac{d}{dz}\psi_{2N-1}
 \end{array}
\right)     \nonumber  \\
+ \frac{w}{2\psi_0}
\left(
 \begin{array}{cccc}
  0 & 0 & \cdots & 0  \\
  \psi_1\psi_0 & \psi_2\psi_0 & \cdots & \psi_N \psi_0  \\
  \vdots & \vdots & \ddots & \vdots  \\ 
  \displaystyle \sum_{k=1}^{N-1}\psi_k \psi_{N-1-k} & 
  \displaystyle \sum_{k=2}^N \psi_k \psi_{N-k} & \cdots & 
  \displaystyle \sum_{k=N}^{2N-2}\psi_k \psi_{2N-2-k}
 \end{array}
\right)     \nonumber  \\
+ \frac{w}{2\psi_0}
\left(
 \begin{array}{cccc}
  \psi_0^2 & \psi_0\psi_1+\psi_1\psi_0 & \cdots 
           & \displaystyle \sum_{k=0}^{N-1}\psi_k\psi_{N-1-k}     \\
  \psi_0\psi_1 & \psi_0\psi_2+\psi_1\psi_1 & \cdots 
               & \displaystyle \sum_{k=0}^{N-1}\psi_k \psi_{N-k}  \\
  \vdots & \vdots & \ddots & \vdots  \\
  \psi_0\psi_{N-1} & \psi_0\psi_N+\psi_1\psi_{N-1} & \cdots 
                   & \displaystyle \sum_{k=0}^{N-1}\psi_k \psi_{2N-2-k}
 \end{array}
\right).    \label{shift3:first}
\end{eqnarray}
Here we note that $\psi_n$'s also satisfy 
\begin{equation}
\hspace*{-20pt}
\frac{d\psi_n}{dz}=
(2z-w)\psi_{n-1}+2(n-1)\psi_{n-2}-\frac{\displaystyle \frac{dw}{dz}-4\alpha}{4\psi_0}
\sum_{k=0}^{n-2}\psi_k\psi_{n-2-k},  \label{cont}
\end{equation}
which is proved by induction from equations (\ref{lin1:p34}), (\ref{lin2:p34})
and (\ref{rec1:p34}). The first term of the right hand side of
equation (\ref{shift3:first}) is rewritten using equation (\ref{cont}) as
\begin{eqnarray}
\fl
(2z-w)
\left(
 \begin{array}{cccc}
  \psi_0     & \psi_1 & \cdots & \psi_{N-1}  \\
  \psi_1     & \psi_2 & \cdots & \psi_N      \\
  \vdots     & \vdots & \ddots & \vdots      \\
  \psi_{N-1} & \psi_N & \cdots & \psi_{2N-2}
 \end{array}
 \right)    \nonumber \\
\fl
+\left(
  \begin{array}{cccc}
   0                & 0                & 0      & 0                 \\
   2\psi_0          & 2\psi_1          & \cdots & 2\psi_{N-1}       \\
   \vdots           & \vdots           & \ddots & \vdots            \\
   2(N-1)\psi_{N-2} & 2(N-1)\psi_{N-1} & \cdots & 2(N-1)\psi_{2N-3}
  \end{array}
  \right)     \nonumber  \\
\fl
+\left(
  \begin{array}{cccc}
   0      & 2\psi_0     & \cdots & 2(N-1)\psi_{N-2}  \\
   0      & 2\psi_1     & \cdots & 2(N-1)\psi_{N-1}  \\
   \vdots & \vdots      & \ddots & \vdots            \\
   0      & 2\psi_{N-1} & \cdots & 2(N-1)\psi_{2N-3}
  \end{array}
  \right)    \nonumber \\
\fl
-\frac{\displaystyle \frac{dw}{dz}-4\alpha}{4\psi_0}
\left(
 \begin{array}{cccc}
  0 & 0 & \cdots & 0  \\
  \psi_0^2 & \psi_1\psi_0 & \cdots & \psi_{N-1} \psi_0  \\
  \vdots & \vdots & \ddots & \vdots  \\
  \displaystyle \sum_{k=0}^{N-2}\psi_k \psi_{N-2-k} & 
  \displaystyle \sum_{k=1}^{N-1}\psi_k \psi_{N-1-k} & \cdots & 
  \displaystyle \sum_{k=N-1}^{2N-3}\psi_k \psi_{2N-3-k}
 \end{array}
\right)     \nonumber  \\
\fl
-\frac{\displaystyle \frac{dw}{dz}-4\alpha}{4\psi_0}
\left(
 \begin{array}{cccc}
  0 & \psi_0^2          & \cdots & \displaystyle \sum_{k=0}^{N-2}\psi_k \psi_{N-2-k} \\
  0 & \psi_0 \psi_1     & \cdots & \displaystyle \sum_{k=0}^{N-2}\psi_k \psi_{N-1-k} \\
  \vdots & \vdots       & \ddots & \vdots  \\
  0 & \psi_0 \psi_{N-1} & \cdots & \displaystyle \sum_{k=0}^{N-2}\psi_k \psi_{2N-3-k} 
 \end{array}
\right).
\end{eqnarray}
Applying the scalar product on these terms, we obtain equation (\ref{shift3}). 
We get equations (\ref{shift4})-(\ref{shift6}) by similar calculations.
%%%%%%%%%%%%%%%%%%%%%%%%%%%%%%%%%%%%%%%%%%%%%%%%%%%%%%%%%%%%%%%%%%%%%%

\end{document}